\documentclass[12pt,a4paper,superscriptaddress]{revtex4}
\usepackage{graphicx}
\usepackage{graphics}
\usepackage{amsmath}
\usepackage{dcolumn}
\usepackage{amssymb}
\usepackage{bm}
\usepackage[latin1]{inputenc}
\newcommand{\be}{\begin{equation}}
\newcommand{\ee}{\end{equation}}
\newcommand{\bea}{\begin{eqnarray}}
\newcommand{\eea}{\end{eqnarray}}

\newcommand{\pa}{\partial}

\def\Ds{D\!\!\!\!/}
\def\Bs{B\!\!\!\!/}

\begin{document}

\title{On one-loop corrections in the Horava-Lifshitz-like QED}

\author{M. Gomes}
\affiliation{Instituto de F\'\i sica, Universidade de S\~ao Paulo\\
Caixa Postal 66318, 05315-970, S\~ao Paulo, SP, Brazil}
\email{mgomes,ajsilva,queiruga@if.usp.br}

\author{T. Mariz}
\affiliation{Instituto de F\'\i sica, Universidade Federal de Alagoas\\ 
57072-270, Macei\'o, Alagoas, Brazil}
\email{tmariz@fis.ufal.br}

\author{J. R. Nascimento}

\affiliation{Departamento de F\'{\i}sica, Universidade Federal da 
Para\'{\i}ba\\
 Caixa Postal 5008, 58051-970, Jo\~ao Pessoa, Para\'{\i}ba, Brazil}
\email{jroberto,petrov@fisica.ufpb.br}

\author{A. Yu. Petrov}

\affiliation{Departamento de F\'{\i}sica, Universidade Federal da 
Para\'{\i}ba\\
 Caixa Postal 5008, 58051-970, Jo\~ao Pessoa, Para\'{\i}ba, Brazil}
\email{jroberto,petrov@fisica.ufpb.br}

\author{J. M. Queiruga}
\affiliation{Instituto de F\'\i sica, Universidade de S\~ao Paulo\\
Caixa Postal 66318, 05315-970, S\~ao Paulo, SP, Brazil}
\email{mgomes,ajsilva,queiruga@fma.if.usp.br}

\author{A. J. da Silva}
\affiliation{Instituto de F\'\i sica, Universidade de S\~ao Paulo\\
Caixa Postal 66318, 05315-970, S\~ao Paulo, SP, Brazil}
\email{mgomes,ajsilva,queiruga@if.usp.br}

\begin{abstract}
We study the one-loop two point functions of the gauge, scalar and spinor fields for a Horava-Lifshitz-like QED with critical exponent $z=2$.  It turns out that, in certain cases, the dynamical restoration of the Lorentz symmetry at low energies can take place. We also analyze the three point vertex function of the gauge and spinor fields and prove that the triangle anomaly identically vanishes in this theory.
\end{abstract}
\maketitle

\section{Introduction}
The Horava-Lifshitz (HL) approach \cite{Lif,Hor}, which is characterized by an  asymmetry between space and time coordinates, has aroused great interest because it provides  an   improvement of the renormalization capabilities of field theories. In this scheme, the equations of motion of relevant models
are invariant under the rescaling $x^i\to bx^i$, $t\to b^zt$, where $z$, the critical exponent, is a number indicating  the ultraviolet behavior of the theory.
This procedure may turn to be essential to enable the  construction of  renormalizable models at scales where quantum gravity aspects cannot be neglected \cite{Hor}.  Different issues related to the HL gravity, including its cosmological features \cite{HorCos}, exact
solutions \cite{Lu}, black holes \cite{BH} were considered in a number of papers.
 However, since the space-time anisotropy breaks Lorentz invariance, to validate a given anisotropic model as  physically consistent it is necessary to prove that  at low energies  the Lorentz symmetry is approximately realized. Some studies suggest that this behavior is better achieved  in infrared stable models.
 
 It is also worth to point out some examples of studies of the perturbative behavior of the HL-like theories. Some facets of the HL generalizations for the gauge and supersymmetric field  theories were presented in \cite{ed}. Renormalizability of the HL-like scalar field theory models has been discussed in detail in \cite{Anselmi}. The Casimir effect for the HL-like scalar field theory has been considered in \cite{ourcas}. In \cite{cpn} and \cite{gomes} the HL modifications of the $CP^{N-1}$ and nonlinear sigma models were respectively studied. Furthermore, the effective potential for various
 HL models was determined in \cite{Liou}.

In this work, we pursue these investigations by  considering a HL generalization of an Abelian gauge theory. 
For the version of the  model containing only scalar and gauge fields, our one-loop calculations, performed at an arbitrary $d$ dimensional space, indicate that the emergence of the Lorentz symmetry at low energies strictly depends on the dimension: it holds for $d=1$ or $2$
but not for $d>3$; for $d=3$ our calculations are inconclusive, as the region where the restoration could take place is  outside from the pertubative setting.    These  observations are based on explicit one-loop calculations of the two and three points vertex functions.
 As we will show,  this pattern has no universal character and does not happen for a similar model of spinor and gauge fields.

 We discuss also  the problem of anomalies, especially the famous Adler-Bell-Jackiw (ABJ) anomaly (triangle anomaly) \cite{ABJ} implying breaking of the chiral symmetry. It is known that  this anomaly causes ambiguities in the theories with ``small'' Lorentz symmetry breaking \cite{JackAmb}. Therefore, it is relevant to verify the presence of such anomaly in the HL-like extension of the QED.

The structure of this work looks as follows. In the section II we present the model, in section III we discuss the one-loop correction to the two point function of the vector field and in the section IV we obtain the one-loop contributions to the two point functions of the scalar and spinor fields. In section V we consider the  three point function of the spinor and gauge field. In section VI we prove the absence of the  triangle anomaly in the $z=2$ QED. Our conclusions are given in the summary.

\section{An HL like Abelian gauge model}

For the sake of  concreteness, we restrict ourselves to the case $z=2$. 
In this case, the Lagrangian  describing  the model we are interested is
\bea
\label{scal}
L&=&\frac{1}{2}F_{0i}F_{0i}+\frac{a_{1}^{2}}{4}F_{ij}\Delta F_{ij}+D_0\phi (D_0\phi)^*-a_{2}^{2}D_iD_j\phi(D_iD_j\phi)^*-m^4\phi\phi^*+
\nonumber\\&+&
\bar{\psi}(i\gamma^0D_0+(ia_{3}\gamma^iD_i)^2-m^2)\psi,
\eea
where $D_{0,i}=\pa_{0,i}-ieA_{0,i}$ is a gauge covariant derivative, with the corresponding gauge transformations being $\phi\to e^{ie\xi}\phi$, $\phi^*\to e^{-ie\xi}\phi^*$, $\psi\to e^{ie\xi}\psi$, $\bar{\psi}\to \bar{\psi}e^{-ie\xi}$, and $A_{0,i}\to A_{0,i}+\pa_{0,i}\xi$. Our metric is $(+-\cdots -)$, and $\Delta$ denotes the $d$-dimensional Laplacian. 

The parameters $a_{i}, \,i=1,2,3$ were introduced to keep track of the contributions associated to the high derivative terms; they are independent but for simplicity we assume that $a=a_{1}=a_{2}=a_{3}$. 
To keep our analysis restricted to the gauge-matter interaction, we do not introduce a self-coupling of the matter fields.

The free propagators for the scalar and fermionic fields are
\bea
<\phi(k)\phi^*(-k)>&=&\frac{i}{k^2_0-(a^{ 2}\vec{k}^4+m^4)+i\epsilon}, \\
\label{freeprop}
<\psi(k)\bar{\psi}(-k)>&=&i\frac{\gamma^0k_0+\omega}{k^2_0-\omega^2+i\epsilon}=\nonumber\\
&=&\frac{P_+}{k_0-\omega+i\epsilon}-\frac{P_-}{k_0+\omega-i\epsilon},
\eea
where $\omega=a\vec{k}^2+m^2$, and $P_{\pm}=\frac{1\pm\gamma_0}{2}$ are orthogonal projectors.
To find the propagator for the vector field, we choose to work in  the Feynman gauge by adding to (\ref{scal})  the gauge fixing Lagrangian \cite{Farias},
\bea
{\cal L}_{gf} = -\frac12[ (-a^{2}\triangle)^{-\frac12}\partial_{0}A_{0}-(-a^{2}\triangle)^{\frac12}\partial_{i}A_{i}]^2,
\eea
yielding to the free propagators  the  forms
\bea 
<A_iA_j>=-\frac{i\delta_{ij}}{k^2_0-a^{ 2}\vec{k}^4+i\epsilon}; \quad\, <A_0A_0>=i\frac{a^{2}\vec{k}^2}{k^2_0-a^2\vec{k}^4+i\epsilon}.
\eea
These expressions will be used to calculate the one-loop contributions in our theory.

\section{One-loop correction to the vector field propagator}

 To study the one-loop correction to the gauge field, let us  first consider the bosonic sector.  It is easy to see that the interaction  vertices are
\bea
&&e^2A_0A_0\phi\phi^*;\quad\, ieA_0(\phi^*\pa_0\phi-\phi\pa_0\phi^*);\\
&&iea^{2}(\pa_iA_j)[\phi\pa_i\pa_j\phi^*-\phi^*\pa_i\pa_j\phi]+2iea^{2}A_i(\pa_j\phi\pa_i\pa_j\phi^*-\pa_j\phi^*\pa_i\pa_j\phi);\nonumber\\
&&e^2a^{2}A_iA_j(\phi\pa_i\pa_j\phi^*+\phi^*\pa_i\pa_j\phi)-e^2(A_i\pa_j\phi+A_j\pa_i\phi+(\pa_iA_j)\phi)
\nonumber\\&\times&
(A_i\pa_j\phi^*+A_j\pa_i\phi^*+(\pa_iA_j)\phi^*),\nonumber
\eea
so that, in the Fourier representation,  they are given by
\bea
\label{vert}
V_3^{(1)}&=&eA_0(p)\phi(k)\phi^*(-p-k)(2k_0+p_0);\\
V_3^{(2)}&=&-e a^{2}A_i(p)\phi(k)\phi^*(-p-k)(2k_j+p_j)[(k_i+p_i)(k_j+p_j)+k_ik_j];\nonumber\\
V_4^{(1)}&=&e^2A_0(p_1)A_0(p_2)\phi(k_1)\phi^*(k_2)(2\pi)^{d+1}\delta(p_1+p_2+k_1+k_2);\nonumber\\
V_4^{(2)}&=&-e^2a^{2}A_i(p_1)A_j(p_2)\phi(k_1)\phi^*(k_2)(2\pi)^{d+1}\delta(p_1+p_2+k_1+k_2)\nonumber\\&\times&
[k_{1i}k_{1j}+k_{2i}k_{2j}-2\delta_{ij}(k_1k_2)-k_{1i}k_{2j}-k_{2i}k_{1j}-k_{1j}p_{2i}-k_{2j}p_{1i}\nonumber\\&-&
\delta_{ij}(k_1p_2)-\delta_{ij}(k_2p_1)-\delta_{ij}(p_1p_2)]\nonumber,
\eea
where $p=(p_0,\,\vec{p})$, and $(pk)=\vec{p}\cdot\vec{k}=p_ik_i$.
At the one-loop order, there are two types of
contributions as indicated in the Fig. 1. There, the wavy line is for the gauge field, and the solid one -- for the scalar field.

\vspace*{2mm}

\begin{figure}[!h]
\begin{center}
\includegraphics[angle=0,scale=1.00]{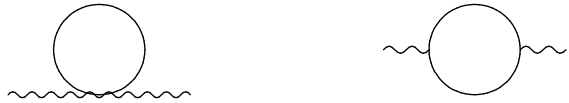}
\end{center}
\caption{Two-point function of the gauge field in the $z=2$ scalar QED}
\end{figure}

\vspace*{2mm}

The tadpole graph gives
\bea
\Pi_0(p) &=& e^2 a^2 A_i(-p) A_j(p)\int\frac{dk_0d^dk}{(2\pi)^{d+1}}\frac{1}{k^2_0-a^{2}\vec{k}^4-m^4} 
[4k_i k_j+2\delta_{ij}\vec{k}^2+\delta_{ij}\vec{p}^{\,\,2}]\label{3}
\nonumber\\
 &&-e^2 A_0(-p) A_0(p) \int\frac{dk_0d^dk}{(2\pi)^{d+1}}\frac{1}{k^2_0-a^{2}\vec{k}^4-m^4},
\eea
where we have omitted the  terms vanishing by symmetry reasons.
Due to the rotational invariance, we can replace the $k_ik_j\to\frac{\delta_{ij}}{d}\vec{k}^2$, so, after integrating in $k_{0}$, we have
\bea
\Pi_0(p)&=&-i\frac{e^2a^{2}A_i(-p)A_i(p)}2\int\frac{d^dk}{(2\pi)^{d}}\frac{1}{\sqrt{a^{2}\vec{k}^4+m^4}}[(\frac{4}{d}+2)\vec{k}^2+\vec{p}^{\,\,2}]\nonumber\\
&&+i\frac{e^2 A_0(-p)A_0(p)}2\int\frac{d^dk}{(2\pi)^{d}}\frac{1}{\sqrt{a^{2}\vec{k}^4+m^4}},\label{1}
\eea
or finally
\bea
\Pi_0(p)&=& -\frac{ie^{2}m^{2}}{(4\pi)^{d/2}}\left(\frac{a^{2}}{ m^{4}}\right)^{1-d/4} \left[\frac{\Gamma[d/4]\Gamma[1/2-d/4]}{4\Gamma[d/2]\sqrt{\pi}} \vec p^{\,\,2}\right.\nonumber\\
&&\left.+(\frac4d+2)\frac{m^{2}}{a}\frac{\Gamma[-d/4]\Gamma[1/2+d/4]}{4\Gamma[d/2]\sqrt{\pi}}\right]A_{i}(p)A_{i}(-p)\nonumber\\
&&+i\frac{e^{2}}{(4\pi)^{d/2}m^2}\left(\frac{a^{2}}{ m^{4}}\right)^{-d/4} \frac{\Gamma[d/4]\Gamma[1/2-d/4]}{{2\Gamma[d/2]\sqrt{\pi}}} A_{0}(p)A_{0}(-p).\label{2}
\eea
It remains to study the "fish" graph which yields three contributions; the  first of them corresponds to two external $A_i$ fields, the second corresponds to one $A_i$ and one $A_0$ fields, and the third  to two $A_0$ fields:
\bea
\Pi_1(p)&=&e^2a^{4}\frac{A_i(p)A_l(-p)}2\int\frac{dk_0d^dk}{(2\pi)^{d+1}}(2k_j+p_j)(2k_m+p_m)
\nonumber\\&\times&
[(k_i+p_i)(k_j+p_j)+k_ik_j]
[(k_l+p_l)(k_m+p_m)+k_lk_m]
\nonumber\\&\times&
\frac{1}{[k^2_0-a^{2}\vec{k}^4-m^4][(k_0+p_0)^2-a^{2}(\vec k+\vec p)^4-m^4]};\label{4}
\eea
\bea
\Pi_2(p)&=&e^2a^{2}A_i(p)A_0(-p)\int\frac{dk_0d^dk}{(2\pi)^{d+1}}(2k_j+p_j)(2k_0+p_0)
\nonumber\\&\times&
[(k_i+p_i)(k_j+p_j)+k_ik_j]
\nonumber\\&\times&
\frac{1}{[k^2_0-a^{2}\vec{k}^4-m^4][(k_0+p_0)^2-a^{2}(\vec k + \vec p)^4-m^4]}.\label{5}
\eea
\bea
\Pi_3(p)&=&e^2\frac{A_0(p)A_0(-p)}2\int\frac{dk_0d^dk}{(2\pi)^{d+1}}(2k_0+p_0)^2
\nonumber\\&\times&
\frac{1}{[k^2_0-a^{2}\vec{k}^4-m^4][(k_0+p_0)^2-a^{2}(\vec k + \vec p)^4-m^4]}.\label{6}
\eea

To investigate the restoration of the Lorentz symmetry we expand the above integrands in Taylor series up to the second order  at $p=0$. For notational simplicity, in the following expressions it should be understood that  only   the integrals (and not  the
fields) are expanded.
\bea
\Pi_{i}(p) \approx \Pi_{i}(0) +\left. \left.\left.\frac{p_{l}p_{j}}2\frac{\partial^{2}\Pi_{i}}{\partial p_{l}\partial p_{j}}\right\vert_{p=0}+\frac{p^{2}_{0}}2\frac{\partial^{2}\Pi_{i}}{\partial p_{0}^{2}}\right\vert_{p=0} + p_{0}p_{l}\frac{\partial^{2}\Pi_{i}}{\partial p_{0}\partial p_{l}}\right\vert_{p=0}.
\eea
It is easily verified that the zeroth order terms, $\Pi_{1}(0)$ and $\Pi_{3}(0)$ are precisely  cancelled by the contributions coming from   (\ref{2}). Actually, the sum of the corresponding integrands in  (\ref{3}), (\ref{4}) and (\ref{6}) is a total derivative vanishing upon integration.
Notice that, 
due to symmetric integration, also $\Pi_{2}(0)=0$. We then proceed by explicitly calculating the second order derivative terms (for simplicity, $k\equiv \vert\vec{k}\vert$): 
\bea
&&\frac{p_ap_b}2\frac{\partial^{2} \Pi_{1}(0)}{\partial p_a\partial p_b}\nonumber\\
&&=e^{2}a^{4}\int\frac{d^{d}k}{(2\pi)^{d}}\left \{\frac{ i k^{4}[-(4+d)m^{8}+a^{2}(16+d)m^{4}k^{4}+2a^{4}dk^{8}]}
{d(2+d)(m^{4}+a^{2}k^{4})^{7/2}} A_{i}\vec{p}^{\,\,2} A_{i}\right.\nonumber\\
&&\left.-\frac{ i k^{4}[(28+12d+d^{2})m^{8}+2 a^{2}(-20+6d+d^{2})m^{4}k^{4}+a^{4}(12+d^{2})k^{8}]}{d(2+d)(m^{4}+a^{2}k^{4})^{7/2}}A_{i}p_{i}p_{l}A_{l}\right \},
\eea
so that, by performing the integral we get
\bea
&&\frac{p_ap_b}2\frac{\partial^{2} \Pi_{1}(0)}{\partial p_a\partial p_b}= \frac{ie^{2}m^{2}}{(4\pi)^{d/2}}\left(\frac{a^{2}}{m^{4}}\right)^{1-d/4}\frac{\Gamma[1/2-d/4]\Gamma[2+d/4]}{3 d \sqrt{\pi}\Gamma[d/2]} A_{i}\vec{p}^{\,\, 2} A_{i}\nonumber\\
&&-\frac{ie^{2}m^{2}}{(4\pi)^{d/2}}\left(\frac{a^{2}}{m^{4}}\right)^{1-d/4}\frac{(10+d)}{12d\sqrt{\pi}}\frac{\Gamma[1/2-d/4] \Gamma[1+d/4]}{\Gamma[d/2]}A_{i}p_{i}p_{l}A_{l}.
\eea

By adding the part coming from the tadpole graph,  we have that  in coordinate space the term
\bea
\frac{ie^{2}m^{2}}{(4\pi)^{d/2}}\left(\frac{a^{2}}{m^{4}}\right)^{1-d/4}\frac{(10+d)}{48\sqrt{\pi}}\frac{\Gamma[1/2-d/4]\Gamma[d/4]}{\Gamma[d/2]}\, F_{ij}F_{ij} 
 \eea 
 will be generated by  the radiative corrections. Besides that,
\bea
\frac{p_{0}^{2}}2\frac{\partial^{2}\Pi_{1}(0)}{\partial p_{0}^{2}}= \frac{ie^{2}}{(4\pi)^{d/2}a} \left(\frac{a^{2}}{m^{4}}\right)^{1-d/4}\frac{\Gamma[1-d/4] \Gamma[3/2+d/4]}{\Gamma[d/2]3d\sqrt{\pi}}A_{i}p_{0}^{2}A_{i},
\eea
\bea
 p_{n}p_{0}\frac{\partial^{2}\Pi_{2}(0)}{\partial p_{0}\partial p_{n}}=-
\frac{2ie^2 }{(4\pi)^{d/2}a}\left(\frac{a^{2}}{m^{4}}\right)^{1-d/4}\frac{\Gamma[1-d/4] \Gamma[3/2+d/4]}{\Gamma[d/2]3 d\sqrt{\pi}}A_{i}p_{0}p_{i} A_{0},
\eea
and,  concerning $\Pi_{3}$, whereas $\frac{\partial^{2}\Pi_{3}(0)}{\partial p_{0}^{2}}=0$, 
\bea
&&\frac{p_ap_b}2\frac{\partial^{2} \Pi_{3}(0)}{\partial p_a\partial p_b}\nonumber\\
&&=\frac{ie^{2}}{(4\pi)^{d/2}a}\left(\frac{a^{2}}{m^{4}}\right)^{1-d/4}\frac{\Gamma[1-d/4] \Gamma[3/2+d/4]}{\Gamma[d/2]3 d\sqrt{\pi}}A_{0}(\vec{p})^{\,\,2}A_{0}.
\eea
The other second order derivatives of $\Pi_{2}$ all vanish due to symmetry reasons. Our results indicate that at low momenta the effective action will be dominated by
\bea
S_{AA}=\int dt\,d^{d}x [\frac12(1+\alpha) F_{0i}F_{0i} -\frac14 \beta m^{2} F_{ij}F_{ij }],
\eea  
 where the factor $\frac12$ comes from the classical action, and  $\alpha$ and $\beta$ are determined from above, being given by
\bea
\label{alfabeta}
\alpha&=&-\frac{2e^{2}}{(4\pi)^{d/2}a} \left(\frac{a^{2}}{m^{4}}\right)^{1-d/4}\frac{\Gamma[1-d/4] \Gamma[3/2+d/4]}{\Gamma[d/2]3d\sqrt{\pi}};\nonumber\\
\beta&=&-\frac{4e^{2}}{(4\pi)^{d/2}}\left(\frac{a^{2}}{m^{4}}\right)^{1-d/4}\frac{(10+d)}{48\sqrt{\pi}}\frac{\Gamma[1/2-d/4]\Gamma[d/4]}{\Gamma[d/2]}.
\eea
  By rescaling  $x,\, t, A_{0} $ and $A_{i}$ by the rules $\pa_0\to a_1\pa_0$, $\pa_i\to a_2\pa_i$, $A_0\to Z_{1}A_0$, $A_i\to Z_2A_i$ with
\bea  
   (1+\alpha) a_1^2=\beta m^{2} a_2^2\label{11}
\eea   
and $a_{1}Z_{2}=a_{2}Z_{1}$ one may check that this low-energy effective action can be rewritten in the standard form 
\bea
S_{AA}=-\frac12(1+\alpha) (a_1Z_2)^{2} \int dt\,d^{d}x F_{\mu\nu}F^{\mu\nu}.
\eea
The above result holds only if $\frac{(\alpha+1)}{\beta}$ is positive which is satisfied if $d=4 +8n + r$ with $n$ a non-negative integer and
$r \in (-2,2 )$.
Therefore, we have showed that within the low-energy limit, and under special relations of parameters of the theory, the usual free Maxwell action is generated. Unlike \cite{cpn}, we have arrived at this result  without any restrictions on the background field.

Now, let us consider the two point function of the vector field in the spinor QED.
Here,  the vertices are
\bea
\label{vertspin}
V_1&=&ie\bar{\psi}\gamma^0A_0\psi,\quad\, V_2=iea^2\bar{\psi}\gamma^i\gamma^j(A_i\pa_j+A_j\pa_i)\psi=2iea^2\bar{\psi}A^i\pa_i\psi,\nonumber\\
V_3&=&iea^2\bar{\psi}\gamma^i\gamma^j(\pa_iA_j)\psi=iea^2\bar{\psi}(\pa_iA^i+\frac{1}{2}\gamma^{ij}F_{ij})\psi, \quad\, V_4=e^2a^2\bar{\psi}A^iA_i\psi.\label{8}
\eea
In the momentum space they look like
\bea
V_1&=&iea^2\bar{\psi}(k)\gamma^0A_0(p)\psi(-p-k),\quad\, V_2=-2ea^2(p_i+k_i)\bar{\psi}(k)A^i(p)\psi(-p-k);\nonumber\\
V_3&=&ea^2p_i\bar{\psi}(k)\gamma^i\gamma^jA_j(p)\psi(-p-k),\nonumber\\ V_4&=&e^2a^2\bar{\psi}(k_1)A^i(p_1)A_i(p_2)\psi(k_2)(2\pi)^{d+1}\delta(k_1+k_2+p_1+p_2).\label{9}
\eea

There are two corresponding Feynman graphs depicted at Fig. 2. There, the dashed line corresponds to fermions. 


\begin{figure}[!h]
\begin{center}
\includegraphics[angle=0,scale=1.00]{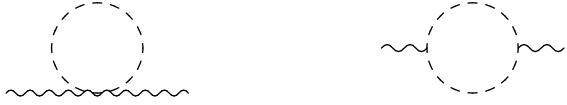}
\end{center}
\caption{Two-point function of the gauge field in the $z=2$ spinor QED}
\end{figure}

\vspace*{2mm}

After a Wick rotation (by the rule $k_0\to ik_{0E}$), the tadpole graph gives the following contribution: 
\bea
\Pi_4(p)&=&-ie^2A_i(-p)A^i(p)\int\frac{dk_{0E}d^dk}{(2\pi)^{d+1}}\frac{{\rm tr}[(\gamma^0k_{0E}+a\vec{k}^2+m^2)]}{k^2_{0E}+(a\vec{k}^2+m^2)^2}.
\eea
By symmetry reasons, and taking into account that ${\rm tr}\,{\bf 1}=D$, with $D$ being the dimension of the gamma matrices, we can rewrite this expression as
\bea
\Pi_4(p)&=&-ie^2DA_i(-p)A^i(p)\int\frac{dk_{0E}d^dk}{(2\pi)^{d+1}}\frac{a\vec{k}^2+m^2}{k^2_{0E}+(a\vec{k}^2+m^2)^2}.
\eea
As $\int\frac{dk_0}{k^2_0+A^2}=\frac{\pi}{A}$, we arrive at the complete cancellation of the factors $a\vec{k}^2+m^2$ in the numerator and in the denominator. Thus,
$\Pi_4(p)\propto \int\frac{d^dk}{(2\pi)^d}$, 
but such  "integral of a constant" vanishes in the dimensional regularization. It remains to analyze the second  graph. Its contribution looks like
\bea
\Pi_5(p)&=&-\frac{e^2}{2}{\rm tr}\int\frac{dk_0d^dk}{(2\pi)^{d+1}}\Big(\gamma^0A_0(-p)+2a^2A_i(-p)k_i+a^2p_i\gamma^i\gamma^jA_j(-p)\Big)\nonumber\\&\times&
\frac{\gamma^0k_0+a\vec{k}^2+m^2}{-k^2_0+(a\vec{k}^2+m^2)^2}
\\&\times&\Big(\gamma^0A_0(p)+2a^2A_l(p)(k_l+p_l)-a^2p_k\gamma^k\gamma^lA_l(p)\Big)
\frac{\gamma^0(k_0+p_0)+a(\vec{k}+\vec{p})^2+m^2}{-(k_0+p_0)^2+(a(\vec{k}+\vec{p})^2+m^2)^2}.\nonumber
\eea
After a long but straightforward calculation, it turns out that both  $F_{0i}F_{0i}$ and $F_{ij}F_{ij}$ parts of this contribution identically vanish in an arbitrary space dimension. 
 Furthermore, we found that in $2+1$ dimensions,
up to one loop order there is no contribution to a Chern-Simons-like term. This result is strictly dependent on the absence of low order spatial derivative terms in the starting Lagrangian while otherwise, the Maxwell (and Chern-Simons) terms are generated.

 It is worth to point out that in a  study  performed in \cite{Iengo} for a HL-like extended spinor QED, in the five-dimensional case,  it  was also obtained  the absence  of one-loop correction to the two point photon function. 
The basic reason why these contributions vanish is the fact that, as it happens  in the  usual nonrelativistic field theory, the contribution from a closed fermionic loop with at least two lines can be decomposed into a sum of two terms each of them having poles only in the same half plane of the  $k_{0}$ variable, see (\ref{freeprop}).  Nonetheless,  our action for the spinor QED differs from that one considered in \cite{Iengo}.

\section{Two point functions of the matter fields}

We can now calculate the two point functions of the matter fields. First, let us consider the scalar QED given by (\ref{scal}). 
The vertices are given again by (\ref{vert}), and the Feynman diagrams are depicted at Fig. 3.

\vspace*{2mm}

\begin{figure}[!ht]
\begin{center}
\includegraphics[angle=0,scale=1.00]{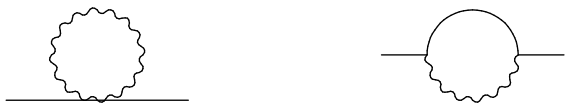}
\end{center}
\caption{Two-point function of the scalar field}
\end{figure}

\vspace*{2mm}

It is clear that within the framework of the dimensional regularization, the tadpole Feynman graph identically vanishes. The fish graph, after some simple arrangements, yields the following contribution from the $<A_iA_j>$ propagator 
\bea
\Xi_1(p)&=&e^2a^4\phi^*(-p)\phi(p)\int\frac{dk_0d^dk}{(2\pi)^{d+1}}\frac{1}{[k^2_0-a^{2}\vec{k}^4][(k_0+p_0)^2-a^{2}(\vec k + \vec p)^4-m^4]}
\nonumber\\&\times&(k_j+2p_j)(k_l+2p_l)[p_ip_j+(k_i+p_i)(k_j+p_j)][p_ip_l+(k_i+p_i)(k_l+p_l)],
\eea
and the following contribution from the $<A_0A_0>$ propagator
\bea
\Xi_2(p)&=&e^2a^4\phi^*(-p)\phi(p)\int\frac{dk_0d^dk}{(2\pi)^{d+1}}\frac{\vec{k}^2}{[k^2_0-a^{2}\vec{k}^4][(k_0+p_0)^2-a^{2}(\vec k + \vec p)^4-m^4]}
\nonumber\\&\times&(2p_0+k_0)(-2p_0-k_0).
\eea
After integration, we arrive at the expression for the small momenta two point function of the scalar field:
\bea
\label{sigmasc}
\Xi&=&\frac{e^2}{2}\int dtd^dx{\phi}^{*}\Big[
-\frac{ 
   \left(a^2 (d+2)+2\right)  \Gamma
   \left(-\frac{d}{4}-1\right)}{2^{\frac{3}{2}d+2} \pi ^{d/2} a^{\frac{d}{2}+1}\Gamma
   \left(\frac{d}{4}\right)}m^d\nonumber\\&-&
\frac{  \left(d \left(a^2
   (d+2)-d-18\right)-128\right)  \Gamma
   \left(-\frac{d}{4}-2\right)}{2^{\frac{3 d}{2}-5} \pi ^{d/2} 
   a^{\frac{d}{2}+1}\Gamma
   \left(\frac{d}{4}\right)}m^{d-4}\pa^2_0\label{7}\\&+&
\frac{ 
   \left(a^2 (d (d (2 d+5)-216)-852)+2 d^2-8\right) 
   \Gamma \left(-\frac{d}{4}-\frac{5}{2}\right)}{2^{\frac{3 d}{2}-7} \pi ^{d/2}  a^{d/2}\Gamma
   \left(\frac{d+2}{4}\right)}m^{d-2}\Delta
\Big]\phi.\nonumber
\eea
By including the  contributions of the tree approximation,  we conclude that for small momenta the scalar sector  of the effective action can be presented as
\bea
S_{\phi^{*}\phi}=\int dt\,d^dx {\phi}^{*}\Big[\alpha_1-m^{4}+(-1+\alpha_2)\pa^2_0+\alpha_3 m^{2}\Delta\Big]\phi,\label{10}
\eea
where $\alpha_1,\alpha_2,\alpha_3$ are  constants whose values can be read from the expression (\ref{sigmasc}). We note that if $d$ is odd, this correction is finite.

We are now in position to discuss the restoration of Lorentz symmetry both in the gauge and scalar sectors. By rescaling (\ref{10}), as it was done after (\ref{alfabeta}), from (\ref{11}) and (\ref{10}) the restoring of the Lorentz symmetry requires that the relations
\bea
 (1+\alpha) a_1^2&=&\beta m^{2} a_2^2, \\
(1-\alpha_{2}) a_{1}^{2} &=& \alpha_{3} m^{2}a_{2}^{2}
\eea
be simultaneously satisfied, which is possible only if 
\bea
(1+\alpha)\alpha_{3}=\beta(1-\alpha_{2}).
\eea
This condition is  an equation for the free  dimensionless parameter $\frac{e^2}{m^{4-d}}$. In fact,
using the explicit values for $\alpha$ and $\beta$ from (\ref{alfabeta}), and
\bea
\alpha_2&=&\frac{e^2}{m^{4-d}}\frac{1}{\Gamma(\frac{d}{4})}2^{4-\frac{3d}{2}}\pi^{-d/2}a^{-d/2-1}\Gamma(-\frac{d}{4}-2)[d(a^2(-d-2)+d+18)+128];\\
\alpha_3&=&\frac{e^2}{m^{4-d}}\frac{1}{\Gamma(\frac{d+2}{4})}2^{6-\frac{3d}{2}}\pi^{-d/2}a^{-d/2}\Gamma(-\frac{d}{4}-\frac{5}{2})
[a^2(d(d(2d+5)-216)-852)+2d^2-8],\nonumber
\eea
we can explicitly find the result for $\frac{e^2}{m^{4-d}}$. In the simplest case $a=1$ which we employ for some calculations in this paper, we have
\bea
\frac{e^2}{m^{4-d}}&=&\frac{2^{d-14}(d+4)(d(d(d-49314)+319580)+2113656)
}{d(d(2d-3)-106)-224}\times\nonumber\\&\times&
\pi^{(d-3)/2}\sin(\frac{\pi d}{2})\Gamma(-\frac{d}{4}-\frac{1}{2})\Gamma(\frac{d}{2}+1)\Gamma(\frac{d}{4})
.
\eea 
We see that for $d=3$, one has $\frac{e^2}{m^{4-d}}=\frac{e^2}{m}\sim 111$, so, it is positive as it must be. However, this number is large, whereas for consistency of the perturbative approach one must have $\frac{e^2}{m^{4-d}}<1$. Therefore, our calculation is inconclusive. One can also verify that for $d=4$ one gets, after the rescaling $\frac{e^2}{m^{4-d}}=0$, and for $d\geq 5$ one has $\frac{e^2}{m^{4-d}}<0$; so, for $d\geq 4$ the restoring of the Lorentz symmetry is impossible, while at $d\leq 3$ one indeed arrives at $\frac{e^2}{m^{4-d}}>0$. We conclude that for $d<3$ the Lorentz symmetry may emerge and  the three-dimensional space is a marginal case so that increasing the dimension makes the situation worse. 

We also  obtained graphics describing $\frac{e^2}{m^{4-d}}$ as a function of $a$. For $d=1$, $d=2$ and $d=3$, respectively,  they are drawn in Figs. $4$, $5$ and $6$. Notice that for the stability of the perturbative series the parameter   $a$ cannot be too small.

\vspace*{2mm}

\begin{figure}[!h]
\begin{center}
\includegraphics[angle=0,scale=1.00]{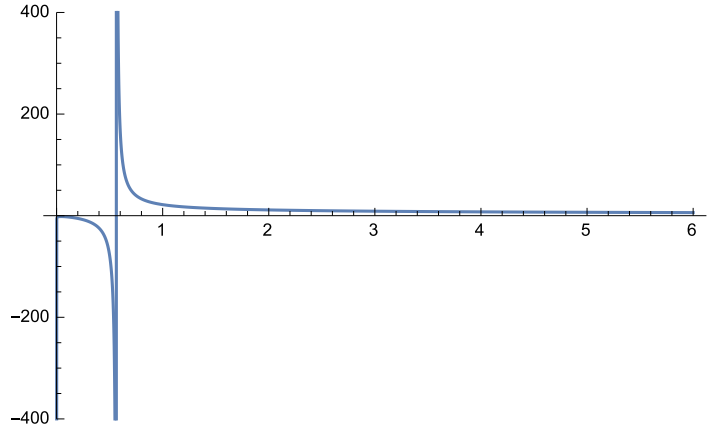}
\end{center}
\caption{The $\frac{e^2}{m^3}$ as a function of $a$, case $d=1$.}
\end{figure}

\vspace*{2mm}

\begin{figure}[!h]
\begin{center}
\includegraphics[angle=0,scale=1.00]{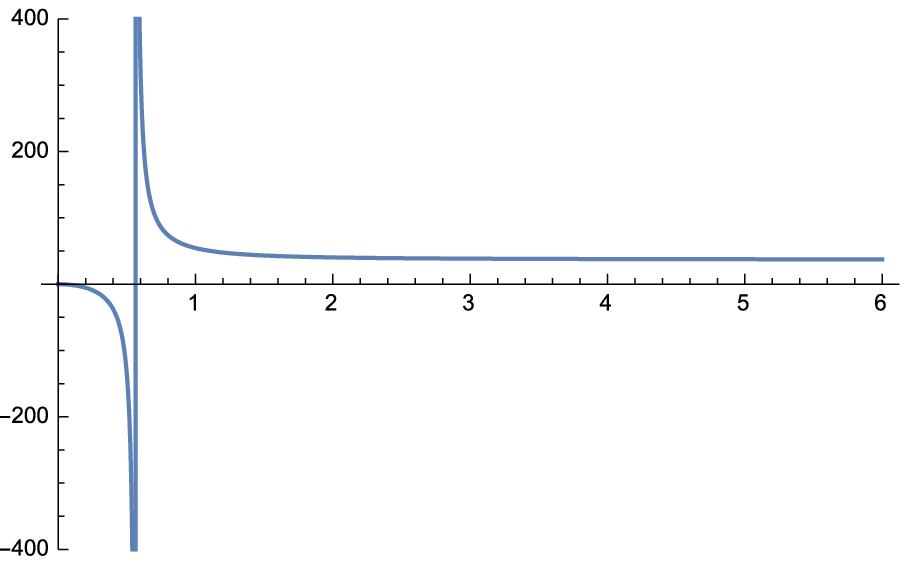}
\end{center}
\caption{The $\frac{e^2}{m^2}$ as a function of $a$, case $d=2$.}
\end{figure}

\vspace*{2mm}

\begin{figure}[!h]
\begin{center}
\includegraphics[angle=0,scale=1.00]{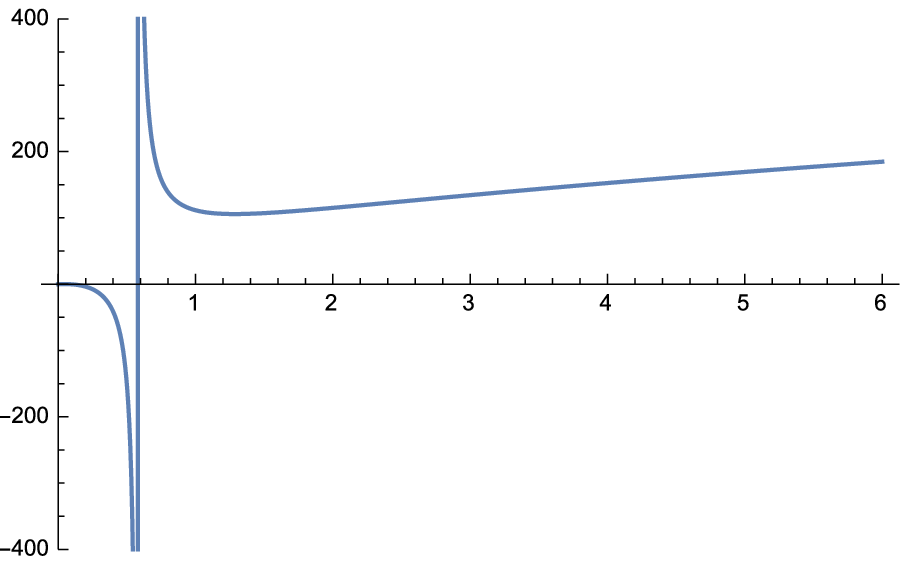}
\end{center}
\caption{The $\frac{e^2}{m}$ as a function of $a$, case $d=3$.}
\end{figure}

For completeness, we give the explicit result for $\frac{e^2}{m^{4-d}}$ for arbitrary $a$ and $d$:
\begin{eqnarray}
&&\frac{e^2}{m^{4-d}} = 2^{d - 11} (d + 4) (d + 8) \pi^{\frac{d - 3}2} a^{\frac d2 - 1} \sin(\frac{\pi d}2) \Gamma(-\frac d4 - \frac12) \Gamma(\frac d2 + 1) \Gamma(\frac d4) \nonumber\\
&&\times\Big(a^2 (d (d ((d-49124) d-122608)+5309456)+20939952)-49152 (d^2-4)\Big)  \nonumber\\
&&\times \Big(3 a^2 (d (d (d (d (5 d+68)-336)-9104)-46096)-72704) \nonumber\\
&&+d (d (d (d (d+60)+1008)+8336)+38192)+74752\Big)^{-1}.
\end{eqnarray}

The spinor QED can be treated in a similar way. In this case, we have the Feynman diagrams depicted in Fig. 7.

\vspace*{2mm}

\begin{figure}[!ht]
\begin{center}
\includegraphics[angle=0,scale=1.00]{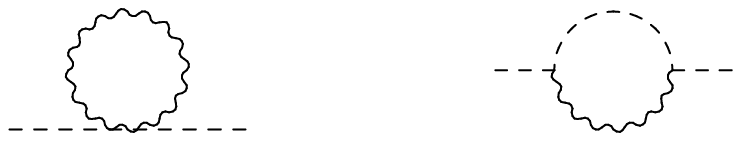}
\end{center}
\caption{Two-point function of the spinor field in the $z=2$ spinor QED}
\end{figure}

\vspace*{2mm}

Again, the tadpole graph  vanishes. The remaining fish graph is obtained by the  contraction of two triple vertices from (\ref{vertspin}). We have
\bea
\Sigma_{\psi}&=&\int \Big(-e^2\bar{\psi}\gamma^0<\psi\bar{\psi}>\gamma^0<A_0A_0>\psi\nonumber\\
&-&e^2a^2\bar{\psi}\gamma^i\gamma^j<\psi\bar{\psi}>\gamma^k\gamma^l\psi<\pa_iA_j\pa_kA_l>\nonumber\\
&-&2e^2a^2\bar{\psi}<\pa_i\psi\bar{\psi}>\gamma^k\gamma^l\psi<A^i\pa_kA_l> - 2e^2a^2\bar{\psi}\gamma^k\gamma^l<\pa_kA_l A^i><\psi\bar{\psi}>\pa_i\psi\nonumber\\&-&
8e^2a^2\bar{\psi}<\pa_i\psi\bar{\psi}>\pa_k\psi<A^iA^k>\Big).
\eea
The explicit form of the one-loop two point function of the spinor field therefore is 
\bea
\Sigma_{\psi}(p)&=&e^2\bar{\psi}(-p)\int\frac{dk_0d^d\vec{k}}{(2\pi)^{d+1}}\Big[-\gamma^0\frac{\gamma^0k_0+a\vec{k}^2+m^2}{k^2_0-(a \vec{k}^2+m^2)^2}\gamma^0\frac{a^2(\vec{p}+\vec{k})^2}{(p_0+k_0)^2-a^2(\vec{p}+\vec{k})^4}\nonumber\\
&+&a^2\gamma^i\gamma^j\frac{\gamma^0k_0+a\vec{k}^2+m^2}{k^2_0-(a \vec{k}^2+m^2)^2}\gamma^k\gamma^l\psi(p)(p_i+k_i)(p_k+k_k)\delta_{jl}\frac{1}{(p_0+k_0)^2-a^2(\vec{p}+\vec{k})^4}\nonumber\\
&+&2a^2\frac{\gamma^0k_0+a\vec{k}^2+m^2}{k^2_0-(a \vec{k}^2+m^2)^2}\gamma^k\gamma^l\psi(p)k^i(p_k+k_k)\delta_{il}\frac{1}{(p_0+k_0)^2-a^2(\vec{p}+\vec{k})^4}\nonumber\\&+&
2a^2p^i(p_k+k_k)\delta_{il}\gamma^k\gamma^l\frac{1}{(p_0+k_0)^2-a^2(\vec{p}+\vec{k})^4}\, \frac{\gamma^0k_0+a\vec{k}^2+m^2}{k^2_0-(a \vec{k}^2+m^2)^2}\nonumber\\&+&
8a^2k_ip_k\delta^{ik}\frac{\gamma^0k_0+a\vec{k}^2+m^2}{k^2_0-(a\vec{k}^2+m^2)^2}\frac{1}{(p_0+k_0)^2-a^2(\vec{p}+\vec{k})^4})\Big]\psi(p).
\eea
The superficial degree of divergence for the fish diagram is $\omega=d-2$. Therefore, for $d=3$, the two point function is linearly divergent. Actually, by symmetry reasons, the divergence is at most logarithmic. 

As before, to verify the possible dynamical restoration of the Lorentz symmetry,  we may explicitly calculate the above integral keeping only the zero and first orders terms in the external momentum $p_0,\,p_i$. 
The result for the general space-time dimension, for $a=1$ is
\bea
\label{ourres}
\Sigma(p)=-ie^2\frac{d-3}{\Gamma(d/2)}m^{d-4}\csc(\frac{d\pi}{2})\pi^{1-d/2}2^{-3d/2}\bar{\psi}\Big[\frac{d-2}{4}\gamma^0\pa_0+\frac{m^2}{2}+O(\partial^2)
\Big]\psi.\label{12}
\eea
In particular, at $d=3$ we arrive at zero result, hence in this case there is no generation of Dirac action and thus no dynamical restoration of the Lorentz symmetry. Also, for any even space dimension more or equal than four, $d=2k\geq 4$, we have a divergent contribution which is in accord with the result found for $d=4$ in \cite{Iengo}.

It is instructive to discuss here  the differences of our results from those obtained in \cite{Iengo}. From a formal viewpoint, the theory (\ref{scal}) considered by us, in the spinor sector, is rather similar to the model discussed in \cite{Iengo}, except for the fact that in our action (\ref{scal}) there is no contributions to the kinetic term corresponding to $z=1$ which however are evidently not relevant in the ultraviolet limit, yielding only subleading terms. First, the two-point function of the photon, in the spinor sector, vanishes both in our case and in the case of $d=4$ discussed in \cite{Iengo} by just the same reasons related to the fact that integrating in the complex plane will allow to close the integration contour with no $k_0$ poles inside.  The spinor propagators in both cases are rather similar, and such that in both cases the integral over $k_0$ vanishes independently of the spatial dimension. However,  the two-point function of the spinor field in our case differs  from that in \cite{Iengo}:  while in \cite{Iengo} it was shown not to be   zero but even divergent, in our case it is zero. There are two differences between our study and that of \cite{Iengo}: first, we use a different gauge (while in \cite{Iengo} the Coulomb gauge was employed, we use a Feynman gauge where the $<A_0A_0> $ propagator is not static), second, which is more important, while in \cite{Iengo} the spatial dimension is fixed in $d=4$, in our case we performed a study in arbitrary dimension with results that agree with \citep{Iengo} in the particular case $d=4$.

\section{The trilinear vertex in the modified spinor QED}

To proceed with our one-loop analysis we will now study low momentum corrections to the  three-point function $\langle\bar{\psi}A\psi\rangle$. That study is based  in the Feynman diagram showed in Fig. 8.

\vspace*{2mm}

\begin{figure}[!ht]
\begin{center}
\includegraphics[angle=0,scale=1.00]{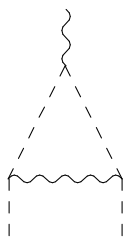}
\end{center}
\caption{Three-point function in the $z=2$ spinor QED}
\end{figure}

\vspace*{2mm}

There are twelve contributions to it. However, as we are mainly interested in the low energy regime, we restrict this study to the usual non derivative terms $\bar{\psi}\gamma^{0,i}\psi A_{0,i}$, and therefore  disregard all dependence on the external momenta. Thus, we have to consider only six following parts:
\bea
T_1(p_1,p_2)&=&e^3\int\frac{d^dkdk_0}{(2\pi)^{d+1}}\bar{\psi}(p_2)\gamma^0G(k)
\gamma^0A_0(p)G(k+p)\gamma^0\psi(p_1)\nonumber\\&\times&<A_0(-k+p_2)A_0(k-p_2)>;\nonumber\\
T_2(p_1,p_2)&=&2a^2e^3\int\frac{d^dkdk_0}{(2\pi)^{d+1}}\bar{\psi}(p_2)\gamma^0G(k)
A^i(p)(k_i+p_i)G(k+p)\gamma^0\psi(p_1)\nonumber\\&\times&
<A_0(-k+p_2)A_0(k-p_2)>;\nonumber\\
T_3(p_1,p_2)&=&-a^4e^3\int\frac{d^dkdk_0}{(2\pi)^{d+1}}\bar{\psi}(p_2)
\gamma^i\gamma^jG(k)\gamma^0
A_0(p)G(k+p)\gamma^k\gamma^l\psi(p_1)\nonumber\\&\times&
(k_i-p_{2i})(k_k-p_{2k})<A_j(-k+p_2)A_l(k-p_2)>;\nonumber\\
T_4(p_1,p_2)&=&-4a^4e^3\int\frac{d^dkdk_0}{(2\pi)^{d+1}}\bar{\psi}(p_2)G(k)
\gamma^0A_0(p)G(k+p)\gamma^i\gamma^j(k-p_2)_i\nonumber\\&\times&<A_j(-k+p_2)A_l(k-p_2)>k^l\psi(p_1);\nonumber
\eea
\bea
T_5(p_1,p_2)&=&-8a^6e^3\int\frac{d^dkdk_0}{(2\pi)^{d+1}}\bar{\psi}(p_2)G(k)
A^l(p)(k_l+p_l)G(k+p)\gamma^i\gamma^j(k-p_2)_i\nonumber\\&\times&<A_j(-k+p_2)A_k(k-p_2)>k^k\psi(p_1);\nonumber\\
T_6(p_1,p_2)&=&-2a^6e^3\int\frac{d^dkdk_0}{(2\pi)^{d+1}}\bar{\psi}(p_2)
\gamma^i\gamma^jG(k)
A^r(p)(k_r+p_r)G(k+p)\nonumber\\&\times&\gamma^k\gamma^l\psi(p_1)
<A_j(-k+p_2)A_l(k-p_2)>(-k_i+p_{2i})(k_l-p_{2l}).
\eea
Here $G(k)=<\psi(-k)\bar{\psi}(k)>$ is the spinor propagator, and $p=-(p_1+p_2)$ is the momentum carried by the external gauge field. 
By setting in the propagators and vertices   the external momenta $p_{1}=p_{2}=p=0$
and taking into account that all propagators are even with respect to $\vec{k}$, we find that only $T_1,T_3$ and $T_4$ survive. We rest with 
\bea
T_1(p_1,p_2)&=&e^3\int\frac{d^dkdk_0}{(2\pi)^{d+1}}\bar{\psi}(p_2)\gamma^0G(k)
\gamma^0A_0(p)G(k)\gamma^0\psi(p_1)\times\nonumber\\&\times&<A_0(-k)A_0(k)>;\nonumber\\
T_3(p_1,p_2)&=&-a^4e^3\int\frac{d^dkdk_0}{(2\pi)^{d+1}}\bar{\psi}(p_2)
\gamma^i\gamma^jG(k)\gamma^0
A_0(p)G(k)\gamma^k\gamma^l\psi(p_1)\times\nonumber\\&\times&
k_ik_k<A_j(-k)A_l(k)>;\nonumber\\
T_4(p_1,p_2)&=&-2a^4e^3\int\frac{d^dkdk_0}{(2\pi)^{d+1}}\bar{\psi}(p_2)G(k)
\gamma^0A_0(p)G(k)\gamma^i\gamma^jk_i\times\nonumber\\&\times&<A_j(-k)A_l(k)>k^l\psi(p_1).
\eea
Now, let us put $a=1$ for simplicity henceforth. Proceeding with calculations in an arbitrary space-time dimension $d$, we obtain the following sum of these three expressions:
\bea
T=-ie^3\frac{(d-3)(d-2)}{\Gamma(d/2)}m^{d-4}\csc(\frac{d\pi}{2})\pi^{1-d/2}2^{-2-3d/2}\bar{\psi}\gamma^0A_0\label{11a}
\psi.
\eea
Therefore, the contribution to the three-point function vanishes for $d=3$ and $d=2$. 
Combining (\ref{11a}) with (\ref{12}) we have
\bea
\Sigma+T=-e^2\frac{(d-3)(d-2)}{\Gamma(d/2)}m^{d-4}\csc(\frac{d\pi}{2})\pi^{1-d/2}2^{-2-3d/2}\bar{\psi}\gamma^0(\partial_{0}+ie A_0)\psi
\eea
as a contribution to the effective Lagrangian, showing that the quantum corrections preserved the gauge symmetry.

\section{Triangle anomaly}

Let us make a last observation about the triangle anomaly in the $z=2$ spinor QED. 
This anomaly appears in the Feynman diagram depicted at Fig. 9, where the upper vertex is for the axial current $\bar{\psi}\Bs\gamma_5\psi$.


\vspace*{2mm}

\begin{figure}[!ht]
\begin{center}
\includegraphics[angle=0,scale=1.00]{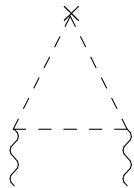}
\end{center}
\caption{Three-point function in the $z=2$ spinor QED}
\end{figure}

\vspace*{2mm}



 Using the same argumentation as in the section IV we can show that the contribution from this graph identically vanishes since again, integrating over $k_0$ in the complex plane, we can close the integration contour with no $k_0$ poles inside.  This is justified by the fact that the projectors $P_{\pm}$ (see (\ref{freeprop})) commute with all combinations of Dirac matrices in the vertices (\ref{vertspin}). It is easy to see that the same argumentation can be applied as well to any even $z$. Therefore, there is no triangle anomaly and no generation of the CFJ term \cite{CFJ}. The less trivial results can be obtained for $z$ odd and at the finite temperature (this study is in progress).
 
The vanishing of the anomaly can also be argued through the path integral formalism. Here we follow the lines presented in \cite{Fuji}.
It is well known that in the case of the usual relativistic QED the chiral transformations of the spinor fields
\bea
\psi\to e^{i\alpha\gamma_5}\psi, \quad \bar{\psi}\to \bar{\psi}e^{i\alpha\gamma_5}
\eea
yield the  contribution to the anomaly caused by the measure transformation and  is equal to
\bea
\delta S_{measure}=\frac{1}{16\pi^2}\int d^4x\alpha(x)\epsilon^{abcd}F_{ab}F_{cd}.
\eea
Let us consider the possibility of the analogous contribution in our case.
Instead of the base of fields $\phi_n$ satisfying the equation $\gamma^aD_a\phi_n=\lambda_n\phi_n$ (remind that the usual spinor action is $\int d^4x\bar{\psi}(i\gamma^aD_a-m)\psi$), in our case, we will consider the base $\tilde{\phi}_n$ satisfying the equation $(i\gamma^0D_0-\Ds^2)\tilde{\phi}_n=\tilde{\lambda}_n\tilde{\phi}_n$, cf. (\ref{scal}).

The corresponding Jacobian can be written as 
\bea
W=\exp[-2i\int dtd^3x\alpha(t,x )A(t,x)],
\eea
where
\bea
A(t,x)= \lim\limits_{N\to\infty}\sum\limits_{l=1}^{N}\tilde{\phi}^{\dagger}_l(t,x)\gamma_5\tilde{\phi}_l(x).
\eea
To provide the convergence of the summation, we introduce a regularization by inserting the function $f(\frac{\tilde{\lambda}^2_l}{\Lambda^4})$, where $\Lambda$ is a constant with a dimension of mass, with $f(x)|_{x\to\infty}=0$ and $f(0)=1$. After proceeding as in \cite{Fuji}, the factor $A$ determining the Jacobian takes the form
\bea
A={\rm Tr}\gamma_5f\left(\frac{(i\gamma^0D_0-\Ds^2)^2}{\Lambda^4}\right),
\eea
where, unlike \cite{Fuji}, $\Ds=\gamma^iD_i$ is a purely spatial contraction. The explicit form of the trace is
\bea
A={\rm tr}\int \frac{d^{4}k}{(2\pi)^{4}}e^{-ikx}\gamma_5f\left(\frac{(i\gamma^0D_0-\Ds^2)^2}{\Lambda^4}\right)e^{ikx},
\eea
where now the trace, $tr$, is only over the gamma matrices. As it is usual in Fujikawa's approach, we may cancel the exponentials in the above expression if at the same time
we replace the covariant derivative $D_{\mu}$ by $D_{\mu}+ i k_{\mu}$.  By expanding the function $f$ in power series of its argument we
may  easily verify that each term either vanishes in the limit $\Lambda\to \infty$ or it is zero because of the trace over the gamma matrices.
For example, the traces of $\gamma_5[\gamma^i,\gamma^j][\gamma^k,\gamma^l]$  and of $\gamma_5\gamma^0[\gamma^i,\gamma^j]$ are  
zero. Hence, the factor $A$ is zero, the Jacobian is trivial and the anomaly  identically vanishes.

\section{Summary}

In this paper we calculated  the low momenta contributions  for the two point function of the gauge field within $z=2$ spinor and scalar QED. Unlike \cite{cpn} where the scalar $z=2$ QED in $2+1$ dimensions has been discussed with a proper time method, we used the Feynman diagram approach. Also, we obtained explicitly the numerical factors accompanying the $F_{0i}F_{0i}$ and $F_{ij}F_{ij}$ contributions. We showed that the Maxwell term naturally emerges. Therefore, at least where the pure gauge sector is concerned, the Lorentz symmetry is dynamically restored. Our studies differ from \cite{cpn} since we, first, considered generic $d$, and second, did not impose any restrictions on the gauge field.

For consistency  and to complete our analysis we studied  the radiative corrections to the two point functions of the matter fields as functions of the spatial dimension $d$. As a result, the restoration of the Lorentz symmetry requires that the dimensionless parameter  $e^{2}/m^{4-d}$  be a precise function of  the parameter $a$ which measures the intensity of the higher spatial derivative term; for the stability of the perturbative series the parameter $a$ cannot be too small. We then found that in the model containing only scalar and gauge fields, the restoration may occur for $d=1$ and $2$ but not for $d>3$. For $d=3$ our result is inconclusive as the possible value of $e^{2}/m^{4-d}$, where the restoration may take place, is outside the perturbative region (for $a=1$ it is equal  to 111).

In the case of spinor QED there is no possibility of the restoration as far as there is no term linear in the spatial derivative in the starting Lagrangian. This is so because, as remarked in \cite{Iengo}, the free spinor propagator is a sum two orthogonal projections, each one having pole either in the upper or in the lower half plane of the integration variable. Thus, there is no purely  fermionic radiative correction to the gauge field propagator, what  therefore breaks the Lorentz invariance at any order of perturbation.

 Besides this, at low momenta,   we calculated the two point function  of the spinor field and also  the gauge-spinor three point functions. Our result shows that in general there is no term  of first order in the spatial derivative and that, for $d=2,3$,  the one loop radiative correction vanishes. Furthermore,  for generic space dimension there are no renormalization of the charge and of the gauge field strength. There is only a wave function renormalization of the spinor field~$\psi$.  

Apparently, to obtain less trivial results in  the fermionic sector,  one should have already at the tree level a term of first order in the spatial derivatives. By a graph analysis and  also by a path integral Fujikawa like method,
we proved  that there is no generation of the ABJ anomaly in our theory.

{\bf Acknowledgements.} This work was partially supported by Conselho
Nacional de Desenvolvimento Cient\'{\i}fico e Tecnol\'{o}gico (CNPq)
and Funda\c{c}\~{a}o de Amparo \`{a} Pesquisa do Estado de S\~{a}o
Paulo (FAPESP). The work by A. Yu. P. has been partially supported by the
CNPq project No. 303438/2012-6. The work of J. M. Q. was supported by
FAPESP.

\newpage

\begin{center}
{\large\bf Erratum: On one-loop corrections in the Horava-Lifshitz-like QED [Phys.Rev.D92:065028,2015]}

{\small M. Gomes, T. Mariz, J. R. Nascimento, A. Yu. Petrov, J. M. Queiruga, A. J. da Silva}

\end{center}

1. In Eq. (16) the right-hand side should be  multiplied  by $(-2)$. 

2. Eq. (17) should be multiplied by $(-1)$.

3. Because of these modifications, the values of the spatial dimension,  quoted  bellow
Eq. (24),  in which a Maxwell action may be  generated, are changed to $d=8n+r$ with $n$ a non-negative integer and $r \in (0,2)$.
This excludes the dimensions $d=2$ and $3$. Notice that the lowest nontrivial value of $d$ is nine but the model is then nonrenormalizable.

4. The factor $a^{4}$ in Eq. (31) should be changed to $a^{2}$. As a consequence, Eq. (32) should be modified as

\setcounter{equation}{31}
\bea
\label{sigmasc}
\Xi&=&\frac{e^2}{2}\int dtd^dx{\phi}^{*}\Big[
\frac{2^{-\frac{3 d}{2}} \pi ^{-\frac{d}{2}}  a^{1-\frac{d}{2}} \Gamma \left(-\frac{d}{4}\right)}{\Gamma \left(\frac{d}{4}\right)}m^d\nonumber\\
&+&\frac{2^{1-\frac{3 d}{2}} \pi ^{-\frac{d}{2}}  a^{1-\frac{d}{2}} \Gamma \left(-\frac{d}{4}-1\right)}{\Gamma \left(\frac{d}{4}\right)}m^{d-4}\partial_0^2\\
&+&\frac{2^{-\frac{3 d}{2}-5} (d (2 d-13)-86) \pi ^{-\frac{d}{2}}  a^{2-\frac{d}{2}} \Gamma \left(-\frac{d}{4}-\frac{3}{2}\right)}{\Gamma \left(\frac{d+2}{4}\right)}m^{d-2}\Delta
\Big]\phi.\nonumber
\eea

5. Because of these changes, Eqs. (37), (38)   and (39) should be replaced by
\setcounter{equation}{36}
\bea
\alpha_2&=&\frac{e^2}{m^{4-d}}\frac{2^{-\frac{3 d}{2}} \pi ^{-\frac{d}{2}}  a^{1-\frac{d}{2}} \Gamma \left(-\frac{d}{4}-1\right)}{\Gamma \left(\frac{d}{4}\right)};\\
\alpha_3&=&\frac{e^2}{m^{4-d}}\frac{2^{-\frac{3 d}{2}-6} (d (2 d-13)-86) \pi ^{-\frac{d}{2}}  a^{2-\frac{d}{2}} \Gamma \left(-\frac{d}{4}-\frac{3}{2}\right)}{\Gamma \left(\frac{d+2}{4}\right)}\nonumber
\eea

and
\setcounter{equation}{38}
\bea
\frac{e^2}{m^{4-d}}=\frac{2^{d+3} (d (2 d (d+15)+223)+498) \pi ^{\frac{d+1}{2}}
   a^{\frac{d}{2}-1} \Gamma \left(\frac{d}{2}\right)}{(d (d (2
   d+27)+374)+1576) \Gamma \left(-\frac{d}{4}-1\right) \Gamma
   \left(\frac{d+6}{4}\right)}
\eea
while the corrected Eq. (38) should be obtained  from the modified Eq. (39) above by setting $a=1$.
 
6. The graphs for $\frac{e^2}{m^{4-d}}$  shown at Figs. 4-6 must be omitted.
 
With these replacements we conclude  that there is no restoration of Lorentz symmetry in the scalar sector. 

None of the conclusions concerning the spinor sector and the anomalies are changed.

\end{document}